\documentclass[%
 reprint,
 amsmath,amssymb,
 aps,
]{revtex4-1}

\usepackage{graphicx}% Include figure files
\usepackage{graphics}
\usepackage{dcolumn}% Align table columns on decimal point
\usepackage{bm}% bold math
\usepackage{epstopdf}
\usepackage{amsmath,amssymb}
\usepackage{float}
\usepackage{epstopdf}
\usepackage{color}
\usepackage{amsfonts}
\usepackage{subfigure}
\usepackage{enumerate}

\begin{document}

\preprint{APS/123-QED}

\title{Bifurcation analysis and potential landscape of the p53-Mdm2 oscillator regulated by the co-activator PDCD5}%

\author{Yuanhong Bi}\email{yuanhong918@163.com}
\affiliation{
School of Mathematics and Systems Science and LMIB, Beihang University, Beijing 100191, China\\
School of Statistics and Mathematics, Inner Mongolia University of Finance and Economics, Hohhot, 010070, China
}

\author{Zhuoqin Yang}\email{yangzhuoqin@buaa.edu.cn}
\affiliation{
School of Mathematics and Systems Science and LMIB, Beihang University, Beijing 100191, China
}

\author{Changjing Zhuge}\email{zhuge@bjfu.edu.cn}
\affiliation{College of Sciences, Beijing Forestry University, Beijing 100083, China
}

\author{Jinzhi Lei}\email{jzlei@tsinghua.edu.cn}
\affiliation{
MOE Key Laboratory of Bioinformatics, Zhou Pei-Yuan Center for Applied Mathematics, Tsinghua University, Beijing 100084, China
}

\date{\today}% It is always \today, today,

\begin{abstract}
Dynamics of p53 is known to play important roles in the regulation of  cell fate decisions in response to various stresses, and PDCD5 functions as a co-activator of p53 to modulate the p53 dynamics. In the present paper, we investigate how p53 dynamics are modulated by PDCD5 during the DNA damage response using methods of bifurcation analysis and potential landscape. Our results reveal that p53 activities can display rich dynamics under different PDCD5 levels, including monostability, bistability with two stable steady states, oscillations, and co-existence of a stable steady state and an oscillatory state. Physical properties of the p53 oscillations are further shown by the potential landscape, in which the potential force attracts the system state to the limit cycle attractor, and the curl flux force  drives the coherent oscillation along the cyclic. We also investigate the effect of PDCD5 efficiency on inducing the p53 oscillations. We show that Hopf bifurcation is induced by increasing the PDCD5 efficiency,  and the system dynamics show clear transition features in both barrier height and energy dissipation when the efficiency is close to the bifurcation point. This study provides a global picture of how PDCD5 regulates p53 dynamics via the interaction with the p53-Mdm2 oscillator and can be helpful in understanding the complicate p53 dynamics in a more complete p53 pathway.
\end{abstract}

% insert suggested PACS numbers in braces on next line% PACS, the Physics and Astronomy
\pacs{Valid PACS appear here}
%\keywords{}%Use showkeys class option if keyword


\maketitle

\section{Introduction}

The tumor suppressor p53 plays a central role in cellular responses to various stress, including oxidative stress, hypoxia, telomere erosion and DNA damage \cite{junttila2009p53, levine2009first}. In unstressed cells, p53 is kept at low level via its negative regulator Mdm2 \cite{kubbutat1997regulation}. Under stressed conditions, such as DNA damage, p53 is stabilized and activated to induce the express of downstream genes, including p21/WAF1/CIP1 and GADD45 that are involved in cell cycle arrest, and PUMA, Bax and PIG3 that can induce apoptosis \cite{hanahan2000hallmarks,hollstein1991p53,laptenko2006transcriptional}. The cell fate decisions after DNA damage are closely related to the p53 dynamics that is regulated by p53-Mdm2 interactions \cite{bar2000generation,GevaZatorsky:2006}. Oscillations of p53 level have been observed upon IR induced DNA damage at the population level in several human cell lines and transgenic mice \cite{bar2000generation,hu2007single,hamstra2006real}. More interestingly, pulses of p53 level were revealed in individual MCF7 cells, and it was suggested that the cell fate is governed by the number of p53 pulses, i.e., few pulses promote cell survival, whereas sustained pulses induce apoptosis \cite{zhang2009cell,zhang2011two}. Fine control of p53 dynamics is crucial for proper cellular response. Mutations and deregulation of p53 expression have been found to associate with various cancer types \cite{brosh2009mutants,puca2010regulation}.

Programmed Cell Death 5 (PDCD5), formerly referred to as TFAR19 (TF-1 cell apoptosis-related gene 19), is known to promote apoptosis in different cell types in response to various stimuli \cite{liu1999tfar19}. Decreased expression of PDCD5 has been detected in various human tumors \cite{spinola2006association,yang2006expression,chen2010prognostic}, and restoration of PDCD5 with recombinant protein or an adenovirus expression vector can significantly sensitive different cancers to chemotherapies \cite{chen2010prognostic,shi2010potent}. PDCD5 is rapidly upregulated after DNA damage, interacts with the apoptosis pathway, and translocates from the cytoplasm to nucleus during the early stages of apoptosis \cite{chen2006short,xu2009pdcd5,chen2001nuclear}. Recently, novel evidence indicates that PDCD5 is a p53 regulator during gene expression and cell cycle \cite{xu2012pdcd5}. It was shown that PDCD5 interacts with p53 by inhibiting the Mdm2-mediated ubiquitination and accelerating the Mdm2 degradation. Hence, upon DNA damage, PDCD5 can function as a co-activator of p53 to regulate cell cycle arrest and apoptosis.

A series of computational models have been constructed to investigate the mechanism of p53-mediated cell-fate decision \cite{bar2000generation,michael2003p53,ma2005plausible,GevaZatorsky:2006,zhang2009cell,zhang2011two}. In these models the p53-Mdm2 oscillation was considered to be crucial for cell-fate decision after DNA damage. After DNA damage, such as double strand breaks (DSBs), active ATM monomer (ATM*) become dominant.  In the nucleus, ATM* active p53 in two ways: ATM* promotes the phosphorylation of p53 at Ser-15 \cite{prives1998signaling} and accelerates the degradation of Mdm2 in nucleus ($\mathrm{Mdm2}_{\mathrm{nuc}}$) \cite{stommel2004accelerated}. In the cytoplasm, active p53 induces the production of $\mathrm{Mdm2}_{\mathrm{cyt}}$, which in turn promotes the translation of \textit{p53} mRNA to form a positive feedback. The p53-Mdm2 oscillation is a consequence of the two coupled feedback loops: the negative-feedback between p53 and $\mathrm{Mdm2}_{\mathrm{nuc}}$, and the positive-feedback between p53 and $\mathrm{Mdm2}_{\mathrm{cyt}}$. Recently, mathematical models of how PDCD5 interacts with the DNA damage response pathway to regulate cell fate decisions have been developed \cite{zhuge2011pdcd5, zhuge2014pdcd5}. In our previous study \cite{zhuge2014pdcd5}, a model for the effect of PDCD5 to the p53 pathway has been established through a nonlinear dynamics model that is developed based on the module of p53-Mdm2 oscillator in \cite{zhang2009cell,zhang2011two} and experimental findings in \cite{xu2012pdcd5}. It was shown that the p53 activity can display different dynamics after DNA damage depending on the PDCD5 level. The p53 protein shows low activity in case of PDCD5 deletion, sustain intermediate level for medial PDCD5 expression, and pulses when the PDCD5 level is upregulated \cite{zhuge2014pdcd5}. Nevertheless, little is known about the global p53 dynamics upon PDCD5 interactions with changes in the expression levels of p53 and PDCD5, which is often seen in tumors.

Here, we investigate, based on the mathematical model proposed in \cite{zhuge2014pdcd5}, how the dynamics of p53 activity after DNA damage depend on changes in the levels of p53 production and PDCD5. A global bifurcation analysis shows that p53 activity can display various dynamics depending on the p53 production and PDCD5 levels, including monostability, bistability with two stable steady states, oscillations, and co-existence of a stable steady state and an oscillatory state. These dynamics are further investigated through the method of potential landscape. The stability of the oscillation states are characterized by the potential force and the probability flux. We also discuss the effect of PDCD5 efficiency on p53 dynamics. PDCD5 efficiency can induce p53 oscillation by Hopf bifurcation, and the transition of p53 dynamics near the Hopf bifurcation are charaterized by barrier height and energy dissipation.

\section{Model and method}

\subsection{PDCD5-mediated p53-Mdm2 oscillator}
\begin{figure}[htbp]
\includegraphics[width=8cm, bb=20 100 700 500]{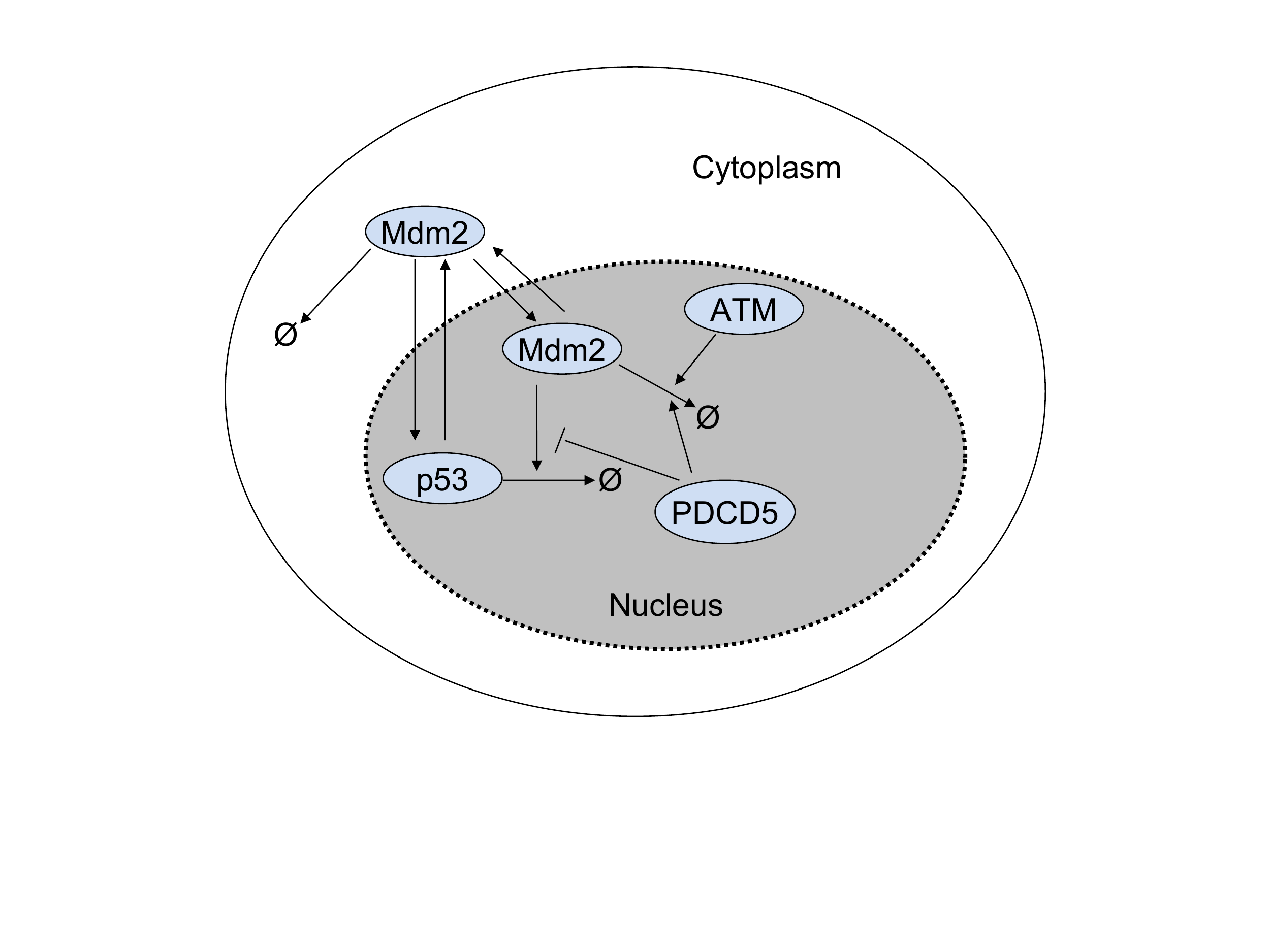}
\caption{Illustration of the PDCD5-mediated p53-Mdm2 pathway. Active p53 promote the production of $\mathrm{Mdm2}_{\mathrm{cyt}}$, which in turn promote the translation of \textit{p53} mRNA. In the nucleus, active p53 is degraded by binding to $\mathrm{Mdm2}_{\mathrm{nuc}}$, and the interaction is disrupted by PDCD5. Both active ATM and PDCD5 are able to accelerate the degradation of $\mathrm{Mdm2}_{\mathrm{nuc}}$. Mdm2 in the nucleus and cytoplasm can shuttle between the two compartments. Refer to the text and \cite{zhuge2014pdcd5} for details.}
\label{fig:1}
\end{figure}

Figure. \ref{fig:1} illustrates the model of p53-Mdm2 oscillator with PDCD5 regulation studied in this paper. Here we summarize the model equations and refer \cite{zhuge2014pdcd5} for details.

The model equations are composed of three components: active p53 in the nucleus $[\mathrm{p53}]$, Mdm2 in the nucleus $[\mathrm{Mdm2}_{\mathrm{nuc}}]$ and in cytoplasm $[\mathrm{Mdm2}_{\mathrm{cyt}}]$. After DNA damage, ATM is activated and promotes the degradation of  $\mathrm{Mdm2}_{\mathrm{nuc}}$ so that the p53 level is unregulated. Active p53 promotes the production of $\mathrm{Mdm2}_{\mathrm{cyt}}$, which promotes the translation of \textit{p53} mRNA to produce p53 and hence form a positive-feedback loop. In the nucleus, active p53 is degraded slowly at a rate $d_{\mathrm{p53}}$ by weakly binding to $\mathrm{Mdm2}_{\mathrm{nuc}}$, and the interaction is disrupted by PDCD5. Moreover, PDCD5 can also accelerate the degradation of $\mathrm{Mdm2}_{\mathrm{nuc}}$. Mdm2 in the nucleus and cytoplasm can shuttle between the two compartments at rates $k_{\mathrm{in}}$ and $k_{\mathrm{out}}$, respectively. These interactions give the following differential equations:
\begin{eqnarray}
\label{eq:p53}
\frac{d[\mathrm{p53}]}{dt}&=&v_{\mathrm{p53}}([\mathrm{Mdm2}_{\mathrm{cyt}}])-d_{\mathrm{p53}}([\mathrm{Mdm2}_{\mathrm{nuc}}])\,[\mathrm{p53}]\nonumber\\
&=& F_{1}([\mathrm{p53}],[\mathrm{Mdm2}_{\mathrm{cyt}}],[\mathrm{Mdm2}_{\mathrm{nuc}}]),    \\
\label{eq:mc}
\frac{d[\mathrm{Mdm2}_{\mathrm{cyt}}]}{dt}&=&v_{\mathrm{Mdm2}}([\mathrm{p53}])-k_{\mathrm{in}}\,[\mathrm{Mdm2}_{\mathrm{cyt}}]\nonumber\\
&&{}+k_{\mathrm{out}}\,[\mathrm{Mdm2}_{\mathrm{nuc}}]-d_{\mathrm{Mdm2}}\,[\mathrm{Mdm2}_{\mathrm{cyt}}]\nonumber\\
&=&F_{2}([\mathrm{p53}],[\mathrm{Mdm2}_{\mathrm{cyt}}],[\mathrm{Mdm2}_{\mathrm{nuc}}]),     \\
\label{eq:mn}
\frac{d[\mathrm{Mdm2}_{\mathrm{nuc}}]}{dt}&=&k_{\mathrm{in}}\,[\mathrm{Mdm2}_{\mathrm{cyt}}]-k_{\mathrm{out}}\,[\mathrm{Mdm2}_{\mathrm{nuc}}]\nonumber\\
&&{}-f(t)\, d_{\mathrm{Mdm2}}\,[\mathrm{Mdm2}_{\mathrm{nuc}}]\nonumber\\
&=&F_{3}([\mathrm{p53}],[\mathrm{Mdm2}_{\mathrm{cyt}}],[\mathrm{Mdm2}_{\mathrm{nuc}}]).
 \end{eqnarray}
Here $k_{\mathrm{in}}, k_{\mathrm{out}}, d_{\mathrm{Mdm2}}$ are constants, and other rate functions are given by
\begin{eqnarray*}
 v_{\mathrm{\mathrm{p53}}}([\mathrm{Mdm2_{cyt}}])&=& \bar{v}_{\mathrm{\mathrm{p53}}}\,\Big[(1-\rho_1) + \rho_1 \,\dfrac{{[\mathrm{Mdm2_{cyt}}]}^{s_1}}{K_1^{s_1}+ {[\mathrm{Mdm2_{cyt}}]}^{s_1}}\Big],\\
d_{\mathrm{p53}}( {[\mathrm{Mdm2_{nuc}}]}) &=& \bar{d}_{\mathrm{p53}}\,\Big[(1-\rho_2)\nonumber\\
&&{}\quad\quad+ \rho_2\,\dfrac{{[\mathrm{Mdm2_{nuc}}]}^{s_2}}{K_2(P(t))^{s_2}+ [\mathrm{Mdm2_{nuc}}]^{{s_2}}}\Big],\\
v_{\mathrm{Mdm2}}({[\mathrm{p53}]}) &=& \bar{v}_{\mathrm{Mdm2}}\,\Big[(1-\rho_3) + \rho_3\,\dfrac{{[\mathrm{p53}]}^{s_3}}{K_3^{s_3}+ {[\mathrm{p53}]}^{s_3}}\Big],\\
f(t) &=& \bar{f}\,\Big[(1-\rho_{4}-\rho_{5}) + \rho_{4}\,\dfrac{P(t)^{s_{4}}}{K_4^{s_{4}}+P(t)^{s_{4}}}\\
&&{} + \rho_{5}\, \dfrac{A(t)^{s_{5}}}{K_5^{s_{5}}+A(t)^{s_{5}}}\Big],
 \end{eqnarray*}
where
\begin{equation}
\label{eq:K2}
K_2(P) = \bar{K}_{2}\,\left((1-r_1) + r_1\,\dfrac{(\alpha_1 P)^{m_1}}{1+(\alpha_1 P)^{m_1}}\right).
\end{equation}
Here $A(t)$ and $P(t)$ are time dependent functions for the levels of active ATM and PDCD5 in nucleus, respectively.

In this study, we only consider the process of DNA repair after DNA damage that can persist for as long as 10 hours before the onset of apoptosis \cite{chen2001nuclear}. Both active ATM and PDCD5 level are upregulated after DNA damage. Hence, we set $A(t) \equiv 5$ to mimic this process in accordance with the simulation given by \cite{zhang2009cell,zhuge2014pdcd5}. Similarly, during DNA repair, we assume $P(t)\equiv P_0$ with $P_0$ represents the PDCD5 expression level.

We refer Table \ref{tab:1} for parameter values in this study.

\begin{table}[t]
 \centering
 \caption{Typical parameter values (The unit for time is \texttt{min}, and arbitrary unit for the concentration)(Ref. \cite{zhuge2014pdcd5}).}
\begin{tabular}{llp{12ex}lp{12ex}lp{12ex}}
\hline\hline&
$	s_{1}	$ 				& $	4	$		&
$	\rho_{1}	$ 			& $	0.991	$	&
$	K_1	$ 					& $	0.057	$	\\ &
$\bar{v}_{\mathrm{p53}}$ 	& $	0.85	$	&
$	s_{2}	$ 				& $	4	$		&
$	\rho_{2}	$			& $	0.9873	$	\\ &
$	\bar{d}_{\mathrm{p53}}$	& $	0.4	$		&
$	m_{1}	$				& $	4	$		&
$	r_{1}	$				& $	0.8	$		\\ &
$	\alpha_{1}	$			& $	3.3	$		&
$	\bar{K}_2	$			& $	0.09	$	&
$	s_3	$					& $	4	$		\\ &
$   \rho_3 $				& $ 0.989$   	&
$   K_3 $					& $ 4.433$		&
$   \bar{v}_{\mathrm{Mdm2}}$& $ 0.135$		\\ &
$	k_\mathrm{in}	$		& $	0.14	$	&
$	k_\mathrm{out}	$		& $	0.01	$	&
$   d_{\mathrm{Mdm2}}$		& $ 0.036	$	\\ &
$   \bar{f} $				& $ 2.7 $		&
$   s_4 $					& $ 4 $			&
$   \rho_4 $ 				& $ 0.2 $		\\ &
$   K_4  $					& $ 0.41 $		&
$   s_5 $					& $ 4 $			&
$   \rho_5 $				& $ 0.5$		\\ &
$   K_5 $					& $1.58$		&\\
\hline\hline\\
\end{tabular}
 \label{tab:1}
\end{table}

\subsection{Potential landscape and probability flux}

The ideas of potential landscape have been introduced for uncovering global principles in biology for protein dynamics \cite{frauenfelder1991energy}, interactions \cite{wolynes1995navigating,wang2003energy}, and gene networks \cite{wang2008potential,wang2010a,wang2010b}. For a non-equilibrium open systems such as the p53-Mdm2 oscillator that exchanges energies and informations with outside environments, the potential and non-equilibrium probability flux are able to reveal insights for the global robustness and physical mechanisms of the non-equilibrium interactions. Here we summarize the potential landscape and flux framework that were introduced by Wang \textit{et al.} in \cite{wang2008potential,wang2010a,wang2010b}.

We write the dynamical equation of the p53-Mdm2 oscillator as $\dot{\mathbf{X}}=\mathbf{F}(\mathbf{X})$, where $\mathbf{X}=([\mathrm{p53}],[\mathrm{Mdm2_{cyt}}],[\mathrm{Mdm2_{nuc}}])$ and $\mathbf{F}(\mathbf{X})$ represents the right hand size of \eqref{eq:p53}-\eqref{eq:mn}. The above equation can be extended to include fluctuations by a probability approach: $\dot{\mathbf{X}}=\mathbf{F}(\mathbf{X}) + \mathbf{\zeta}$, where $\mathbf{\zeta}$ is the noise perturbation. The statistical nature of the noise is often assumed as Gaussian (large number theorem) and white (no memory): $\langle\mathbf{\zeta}(t)\mathbf{\zeta}(t')\rangle = 2\textbf{D}\delta(t-t')$ and $\langle\mathbf{\zeta}(t)\rangle=\mathbf{0}$, where $\mathbf{D}$ is the correlation tensor (matrix) measuring the noise strength.

The probability of system states $P(\mathbf{X},t)$ evolves following the Fokker-Plank equation $\frac{\partial P}{\partial t} + \nabla \cdot \mathbf{J}(\mathbf{X},t)=0$, where the flux vector $\mathbf{J}$ is defined as $\mathbf{J}(\mathbf{X},t) = \mathbf{F} P - \mathbf{D}\cdot \frac{\partial\ }{\partial \mathbf{X}}P$. The flux $\mathbf{J}$ measures the speed of the flow in the concentration space.

At the stationary state, $\frac{\partial P}{\partial t}=0$, then $\nabla \cdot\textbf{J}(\textbf{X},t)=0$. There are two possibilities: one is $\mathbf{J}=0$, which implies detailed balance, and the other is $\mathbf{J}\neq\mathbf{0}$ so that the detailed balance is broken and the system is at the non-equilibrium state.

At the stationary state, the probability flux is defined by 
 \begin{equation}
 \label{eq:Jss}
  \textbf{ J}_{ss}=\textbf{F}P_{ss}-\textbf{D}\cdot \frac{\partial\ }{\partial \textbf{X}}P_{ss}
 \end{equation}
 where $P_{ss}$ is the probability density at stationary state. Hence, 
\begin{eqnarray}
\label{eq:F}
  \mathbf{F}&=&\mathbf{D}\cdot \frac{\partial\ }{\partial \mathbf{X}}P_{ss}/P_{ss}+\mathbf{J}_{ss}/P_{ss}\nonumber\\
  &=&-\mathbf{D}\cdot\frac{\partial\ }{\partial \mathbf{X}}(-\ln P_{ss})+\textbf{J}_{ss}/P_{ss}\nonumber\\
  &=&-\mathbf{D}\cdot\frac{\partial\ }{\partial \mathbf{X}}U+\mathbf{J}_{ss}/P_{ss}.
\end{eqnarray}
Here $U=-\ln P_{ss}$ is defined as the non-equilibrium potential. From \eqref{eq:F}, we have divided the force $\mathbf{F}$ into two parts: the potential force $-\mathbf{D}\cdot \frac{\partial\ }{\partial \mathbf{X}}U$ and curl flux force $\mathbf{J}_{ss}/P_{ss}$. At detail balance, the curl flux force is zero. For a non-equilibrium open system, both potential landscape and the associated flux are essential in characterizing the global stationary state properties and the system dynamics.

\section{Results}

\subsection {Codimension-two bifurcation analysis}

\begin{figure*}[htbp]
\centering
\includegraphics[width=12cm]{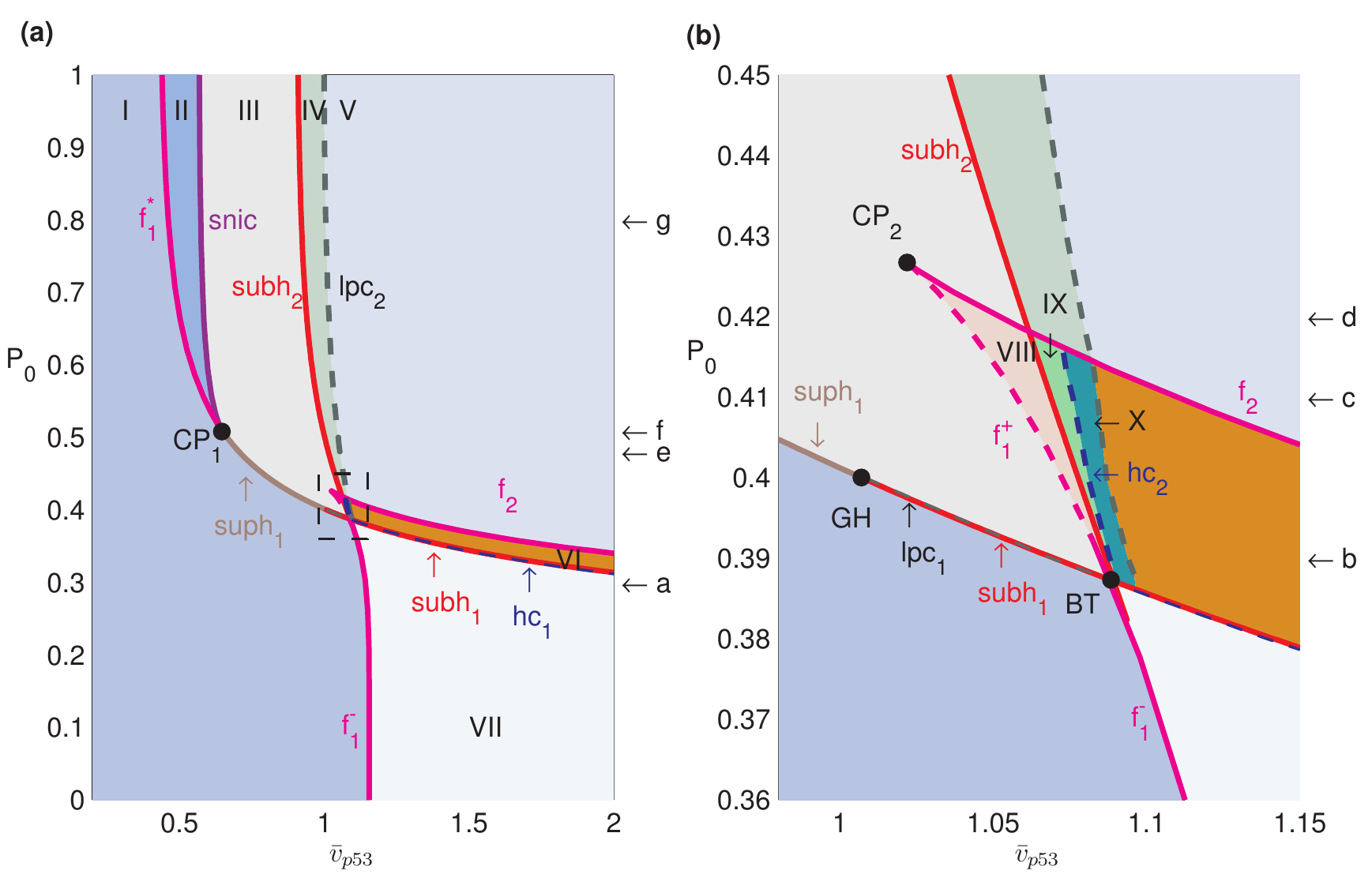}
\caption{Codimension-Two bifurcation diagram. (a) The two-parameter bifurcation diagram with respect to $\bar{v}_{p53}$ and $P_{0}$. (b) Enlarged representation of the rectangle in (a). Codimension-one bifurcation curves are: $suph$--the supercritical Hopf bifurcation; $subh_{1},subh_{2}$-- the subscribe hopf bifurcation; $f_{i}$--the fold bifurcation of equilibria; $lpc_{1},lpc_{2}$--fold bifurcation of limit cycles; $hc_{1},hc_{2}$--homoclinic bifurcation. These curves mainly divide  the ($\bar{v}_{p53}$,$P_{0}$) plane  into ten regions I-X . Codimension-two bifurcation points are: GH-generalized Hopf bifurcation, CP--the cusp bifurcation and BT--Bogdanov-Takens bifurcation. Values of $P_{0}$ denoted by a-g correspond to $P_{0}$ values in Fig. \ref{fig:5}, respectively.}
\label{fig:bif2}
\end{figure*}

To investigate how PDCD5 interacts with p53 to regulate cell fate decision dynamics after DNA damage, we performed codimension-two bifurcation analysis with respect to the two parameters $P_{0}$ and $\bar{v}_{\mathrm{p53}}$. The bifurcation diagrams were computed with AUTO incorporated in XPPAUT \cite{ermentrout2002simulating}. The main bifurcation diagram is shown at Fig. \ref{fig:bif2} and detailed below, with Fig. \ref{fig:bif2}b the enlarge of the dashed square region in Fig. \ref{fig:bif2}a.

\begin{figure*}[htbp]
\centering
\includegraphics[width=15cm]{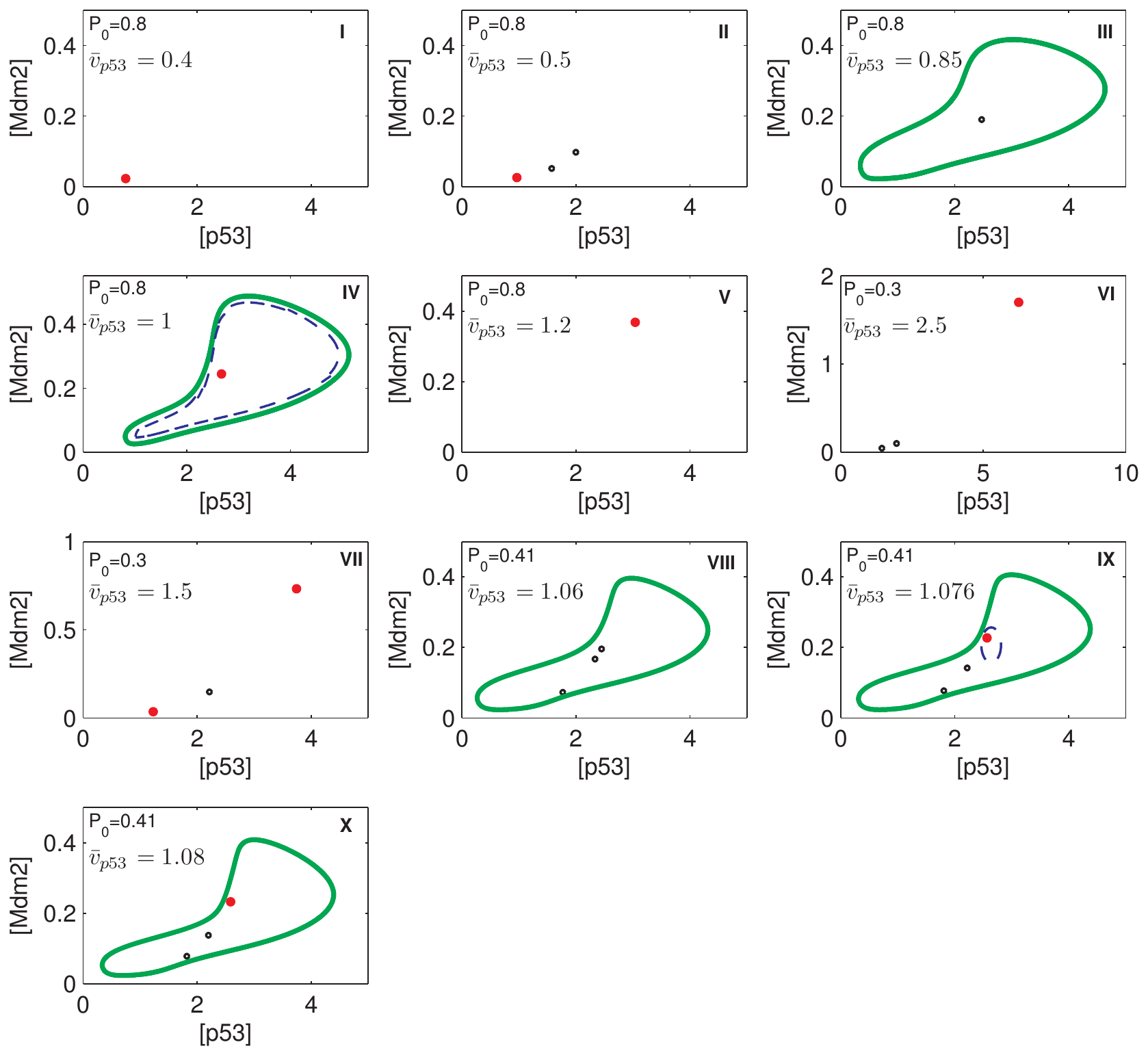}
\caption{Phase diagrams corresponding to the labeled regions I-X in Fig. \ref{fig:bif2}. Red solid dots represent stable equilibria, black open circles unstable one; green solid and blue dash dotted lines represent stable limit cycles and unstable limit cycles respectively. }
\label{fig:dyn}
\end{figure*}

The parameter plane $(\bar{v}_{\mathrm{p53}}, P_{0})$ is divided into ten regions labelled \textit{I}--\textit{X} each with different dynamical profiles and marked with different colors  in Fig. \ref{fig:bif2} (refer Fig. \ref{fig:bif2}b for regions $VIII$-$X$). Typical dynamics of each region is shown at Fig. \ref{fig:dyn} and described below:
\begin{enumerate}
\item Region \textit{I} corresponds to monostability with a single stable steady state of low p53 activity despite the expression level of PDCD5.
\item In region \textit{II} there is a stable steady state and two unstable steady states. Regions \textit{I} and \textit{II} are separated by a curve $f^{*}_1$ of saddle-node bifurcation, across which an unstable node and a saddle appear due to fold bifurcation.
\item Region \textit{III} is on the right of region \textit{II}. Crossing the curve $snic$ from region \textit{II}, the stable node and the saddle collide and disappear, accompanied by the emergence of a stable limit cycle that surrounds an unstable steady state. Meanwhile, parameters can also change from region  \textit{I} to \textit{III} crossing the boundary $suph_1$ or $subh_1$, respectively (Fig. \ref{fig:bif2}b). The curve $suph_1$ represents supercritical Hopf bifurcation at which the stable steady state in region \textit{I} becomes unstable and a stable limit cycle appears. The curve $subh_1$ represents subcritical Hopf bifurcation. Before crossing the curve $subh_{1}$, two limit cycles, stable and unstable ones, come out due to fold bifurcation of limit cycles that is represented by the curve $lpc_{1}$ (Fig. \ref{fig:bif2}b). Next, the unstable limit cycle disappear and the stable steady state lose its stability when at the curve $subh_{1}$.
\item Regions \textit{III} and \textit{IV} are separated by a subcritical Hopf bifurcation curve $subh_{2}$ by which an unstable limit cycle appears and locates between the stable limit cycle and the steady state, and the steady state that is unstable at region \textit{III} becomes stable at region \textit{IV}.
\item Region \textit{V}, similar to region \textit{I}, has a stable steady state which corresponds to a state of high p53 concentration. The two limit cycles at region \textit{IV} collide and disappear when parameters cross the fold bifurcation curve $lpc_{2}$ that separates regions \textit{IV} and \textit{V}.
\item Three steady states arise in region \textit{VI}, two of them are unstable and appear while crossing the fold bifurcation curve $f_{2}$ of equilibria between regions \textit{VI} and \textit{V}.
\item Region \textit{VII} gives bistability in p53 activity, including two stable steady states with low or high p53 concentrations, and an unstable steady state. There are two ways to reach this bistable region. From region \textit{VI} and crossing the subcritical Hopf bifurcation $subh_{1}$, an unstable limit cycle emerges and the unstable steady state with low p53 level becomes stable, but the unstable limit cycle immediately disappear due to homoclinic bifurcation (refer the curve $hc_{1}$ in Fig. \ref{fig:bif2}a which is very close to $subh_{1}$). From region \textit{I} and crossing the boundary $f_{1}^{-}$, a pair of stable and unstable steady states appear due to fold bifurcation of equilibria.
\item There are three regions (\textit{VIII}-\textit{X}) around $(\bar{v}_{\mathrm{p53}}, P_{0})=(1.06,0.41)$ which share the same phase diagram with a stable limit cycle and three steady states (Fig. \ref{fig:bif2}b). In region \textit{VIII} the three steady states are unstable, two of them (a saddle and an unstable focus) come from the fold bifurcation ($f_{1}^{+}$) that separates regions \textit{III} and \textit{VIII}. In region \textit{IX} the steady state with high p53 level becomes stable and is surrounded by an unstable limit cycle. The change is originated from the subcritical Hopf bifurcation $subh_{2}$ from region \textit{VIII}. Finally, the unstable limit cycle disappear in region \textit{X} due to the homoclinic bifurcation $hc_{2}$ between regions \textit{IX} and \textit{X}. We note that in regions \textit{IX} and \textit{X}, there are bistable states of a stable steady state and a stable limit cycle, similar to the region \textit{IV}.
\end{enumerate}

In addition to the above ten regions, there are four codimension-two bifurcation points denoted by black dots in Fig. \ref{fig:bif2}: two \textit{cusp} points ($\mathrm{CP}_{1}$ and $\mathrm{CP}_{2}$), one Generalized Hopf bifurcation (GH) and one Bogdanov-Takens bifurcation (BT). The \textit{cusp} point $\mathrm{CP}_{1}$ locates at $(\bar{v}_{\mathrm{p53}} , P_{0})=(0.6464,0.508)$ where the fold bifurcation curve  $f^{*}_{1}$ and the saddle-node homoclinic bifurcation curve $snic$ meet tangentially. The other \textit{cusp} point $\mathrm{CP}_{2}$ is given by $(1.0220, 0.4267)$ by the two fold bifurcation curves $f_{1}$ and $f_{2}$, where two fold bifurcations coalesce and disappear. The point GH at $(1.0071, 0.4)$ corresponds to the meeting point of $suph_{1}$ and $subh_{1}$ from where a fold bifurcation curve of the limit cycle ($lpc_{1}$) occurs. At the BT point $(1.0885,0.3873)$ the fold bifurcation curve $f_{1}$ and $subh_{2}$ meet tangentially to give a homoclinic bifurcation curve $hc_{2}$. The BT point separates $f_{1}$ into two segments $f_{1}^{-}$ and $f_{1}^{+}$. Crossing $f_{1}^{-}$ from left to right gives a stable equilibrium and a saddle, while crossing $f_{1}^{+}$ gives an unstable equilibrium and a saddle \cite{kuznetsov1998}.

In summary, by manipulating the expression level of PDCD5 and the maximum production rate of p53, the system displays four types of stable dynamics: a single stable steady state (regions \textit{I}, \textit{II}, \textit{V}, and \textit{VI}), two stable steady states (region \textit{VII}), a stable limit cycle (regions \textit{III} and \textit{VIII}), and coexistence of a stable steady state and a stable limit cycle (regions \textit{IV}, \textit{IX}, and \textit{X}). The bifurcation diagram is zoomed out by codimension-one bifurcations with seven $P_{0}$ values marked by $a$-$g$ in Fig. \ref{fig:bif2} and are detailed at the next section.

\begin{figure}
\centering
\includegraphics[width=9cm]{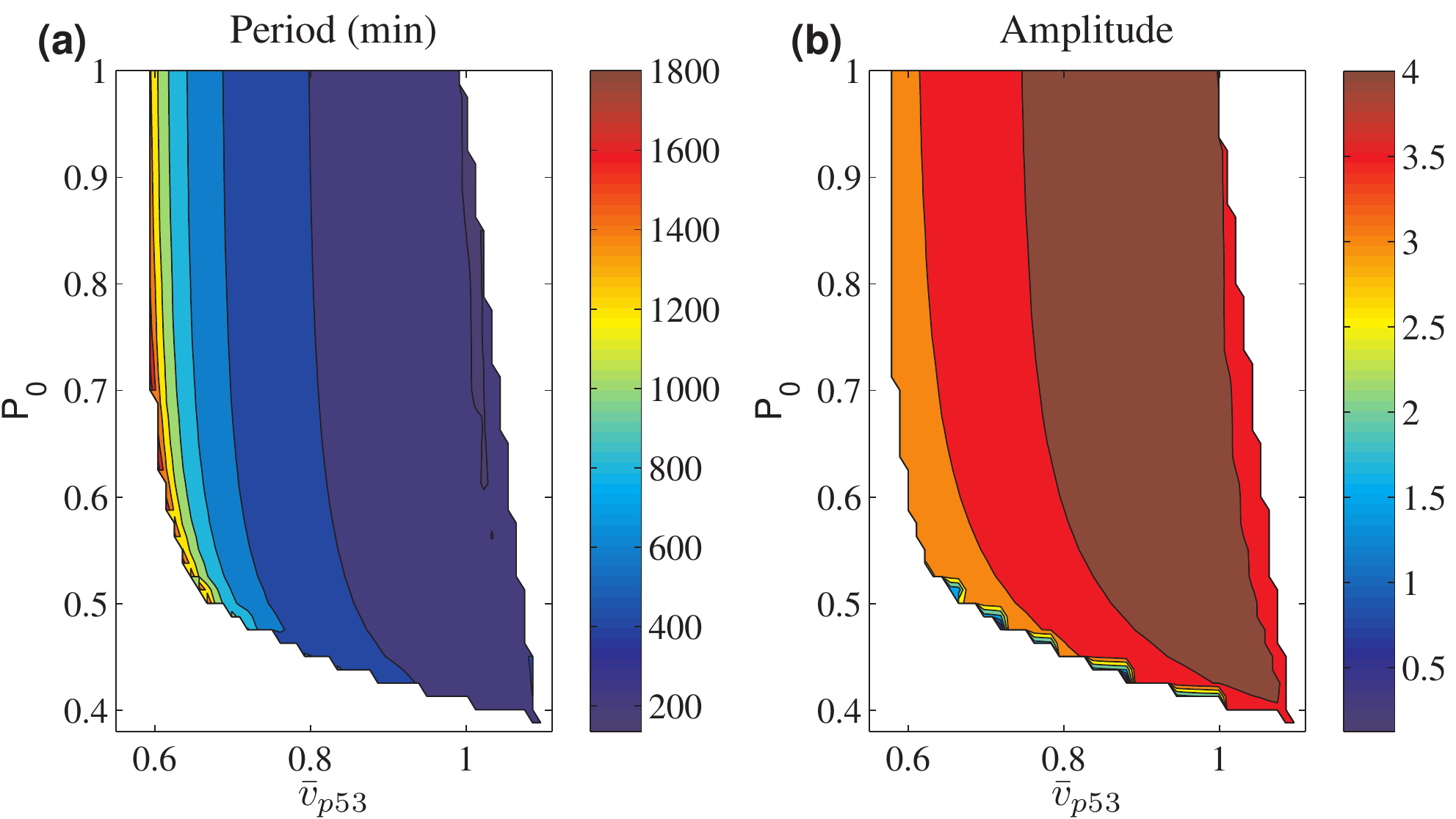}
\caption{Periods (a) and amplitudes (b) of oscillations corresponding to the regions of oscillation dynamics in Fig. \ref{fig:bif2}. }
\label{fig:4}
\end{figure}

From Fig. \ref{fig:bif2}, during DNA damage and when the PDCD5 is upregulated ($P_{0} > 0.3873$), the system shows either low p53 activity, p53 oscillation, or sustained high p53 level with the increasing of p53 production rate. p53 oscillations have been known to be essential for cellular response to DNA damage including DNA repair and apoptosis \cite{Gajjar:2012,zhang2009cell,GevaZatorsky:2006,purvis2012p53}. Our analyses suggest that proper cell response to DNA damage with p53 oscillation is possible only when PDCD5 is highly expressed and the p53 production rate takes proper value. To further investigate the oscillation dynamics, Fig. \ref{fig:4} shows the period and amplitude of the stable limit cycles corresponding to regions \textit{III} and \textit{IV} in Fig. \ref{fig:bif2}. The results show that the oscillation periods decrease with the maximum p53 production rate ($\bar{v}_{\mathrm{p53}}$), while the amplitudes increase with $\bar{v}_{\mathrm{p53}}$. This is consistent with the experimental observations \cite{proctor2008explaining}. However both periods and amplitudes are insensitive with the PDCD5 level over a wide parameter range. These results indicates that upregulated PDCD5 and a proper p53 expression level are essential for producing proper p53 oscillation that regulates the cellular response to DNA damage.

When PDCD5 is downregulated ($P_{0} < 0.3873$), the system has either a single steady state with low p53 activity or bistability with either low or high p53 activity after DNA damage depending on the maximum p53 production rate. Experiments have shown that cells exposed to sustained p53 signaling frequently underwent senescence \cite{purvis2012p53}.

\subsection{Codimension-one bifurcation analysis}

\begin{figure*}[htbp]
\centering
\includegraphics[width=15cm]{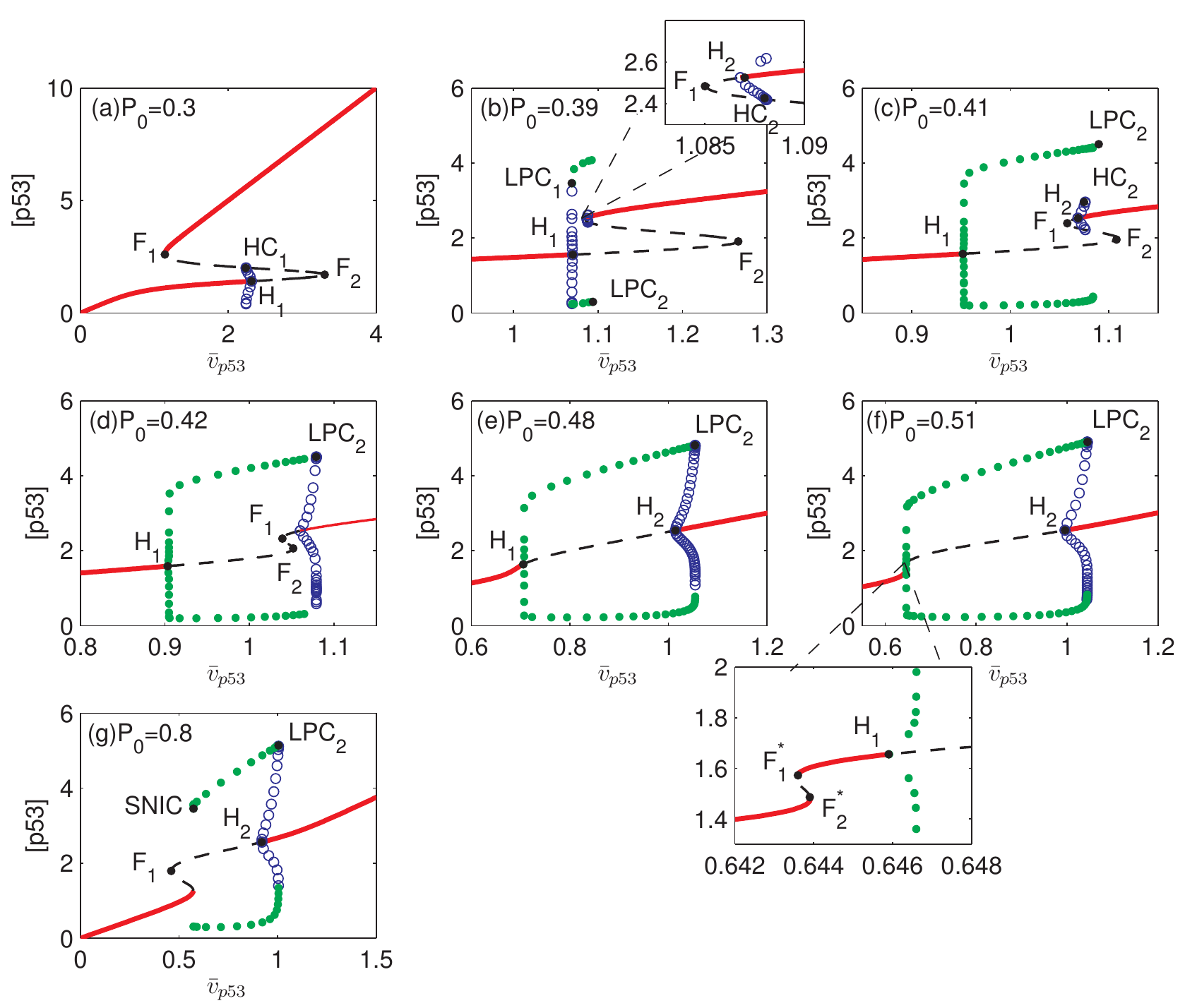}
\caption{Typical codimension-one bifurcation diagrams of [p53] with respect to the parameter $\bar{v}_{p53}$ with given PDCD5 level as labelled by a-g at Figure \ref{fig:bif2}. In all diagrams, red solid lines represent stable equilibria, black dashed lines are unstable equilibria, green solid dots are the maxima and minima of the stable limit cycles, while blue open circles represent the maxima and minima of unstable limit cycles. Codimension-one bifurcation point are marked as $ F_{i}$ (or $F^*_i$) for fold bifurcation points of equilibria, $H_{i}$ for Hopf bifurcation points, $LPC_{i}$ for the fold bifurcation points of limit cycle, \textit{HC} for homoclinic bifurcation points, and \textit{SNIC} for a saddle-node homoclinic bifurcation point.}
\label{fig:5}
\end{figure*}

To get a clear insight into the codimension-two bifurcation diagram in Fig. \ref{fig:bif2}, we considered a codimension-one bifurcation of the concentration of p53 with respect to the p53 production rate $\bar{v}_{p53}$.  The bifurcation diagram is shown at Fig. \ref{fig:5},  with panels a-g for given values $P_0 = 0.3, 0.39, 0.41, 0.42, 0.48, 0.51, 0.8$, respectively, as marked at Fig. \ref{fig:bif2}.  Diagrams for each $P_0$ value is detailed below.
\begin{enumerate}[(a).]
\item For low PDCD5 level ($P_0=0.3$), the equilibrium show a S-shaped bifurcation diagram with bistability (Fig. \ref{fig:5}a). There are three branches of the equilibrium depending on $\bar{v}_{\mathrm{p53}}$. The upper branch is composed of stable nodes, states at the middle branch are saddles, while the lower branch consists of stable nodes and foci which are separated near the subcritical Hopf bifurcation point $H_{1}$ ($\bar{v}_{p53}=2.315$). There is a fold bifurcation of equilibria $F_{1}$ ($\bar{v}_{\mathrm{p53}}=1.143$) where a stable node and a saddle appear, so that bistability occurs with $\bar{v}_{\mathrm{p53}}$ between $F_1$ and $H_1$. With the increasing of $\bar{v}_{p53}$, an unstable limit cycle appears from a homoclinic bifurcation point $HC_{1}$ at a  saddle ($\bar{v}_{\mathrm{p53}}=2.239$), and the unstable limit cycle disappears via the subcritical Hopf bifurcation $H_{1}$. Crossing $H_1$, stable focus at the lower branch change to unstable one. The saddle and the lower unstable focus collide and disappear via another fold bifurcation $F_{2}$ at $\bar{v}_{p53}=3.305$.
\item When $P_{0}$ increases ($P_0 = 0.39$) (Fig. \ref{fig:5}b), the fold bifurcation $F_{1}$ becomes larger than the subcritical Hopf bifurcation $H_{1}$ so that the bistability of equilibrium vanishes.  With the increasing of $\bar{v}_{\mathrm{p53}}$, an unstable limit cycle arises from the Hopf bifurcation $H_1$. The limit cycle becomes stable at  the fold bifurcation $LPC_1$, and then vanishes at the second fold bifurcation $LPC_2$. Different from the case $P_0 = 0.3$, now the fold bifurcation $F_1$ gives a saddle (the middle branch) and an unstable focus (the upper branch) (Fig. \ref{fig:5}b inset).  The unstable focus on the upper branch becomes stable and an unstable limit cycle appears via the subcritical Hop bifurcation $H_{2}$. The unstable limit cycle  develops until it meet the saddle at the homoclinic bifurcation point $HC_{2}$. The stable focus and stable limit cycle coexist with $\bar{v}_{\mathrm{p53}}$ between $H_2$ and $LPC_2$, which gives the region \textit{X} in Fig. \ref{fig:bif2}.
\item When $P_{0}$ increases to $0.41$ (Fig. \ref{fig:5}c), the diagram is similar but the Hopf bifurcation $H_1$ changes from subcritical to supercritical, so that stable limit cycles exist with $\bar{v}_{\mathrm{p53}}$ taken values over a wide range from $H_1$ to $LPC_2$.
\item When $P_{0}=0.42$, $F_{1}$ becomes too close to  $F_{2}$ so that the unstable limit cycle arise from $H_{2}$ does not collide with a saddle but ends up at the fold bifurcation point of limit cycles $LPC_{2}$ (Fig. \ref{fig:5}d).
\item For a larger $P_{0}$ (Fig. \ref{fig:5}e), the two fold bifurcation points $F_{1}$ and $F_{2}$  coalesce and disappear due to a codimension-two \textit{cusp} point $CP_{2}$ in Fig. \ref{fig:bif2}.
\item With the further increasing of $P_{0}$ (Fig. \ref{fig:5}f), another two fold bifurcation points ($F^{*}_{1}$ and $F^*_{2}$) arise from the Hopf bifurcation $H_{1}$ due to the \textit{cusp} point $CP_{1}$ in Fig. \ref{fig:bif2}.
\item Further increasing $P_{0}$ to higher level ($P_0 = 0.8$), the supercritical Hopf bifurcation point $H_{1}$ moves to the left and collides with $F^*_{2}$ to give a fold-Hopf bifurcation point (\textit{ZH}), and then $F^*_{2}$ changes to a saddle-node homoclinic bifurcation point \textit{SNIC}, from which the saddle and the node collide and disappear to generate a stable limit cycle.
\end{enumerate}

The above bifurcation analyses present  the global dynamics of the whole system, which is further explored below through the underlying potential landscape \cite{frauenfelder1991energy,wolynes1995navigating}.

\subsection { Potential landscapes and global dynamics}

\begin{figure*}[htbp]
\centering
\includegraphics[width=15cm]{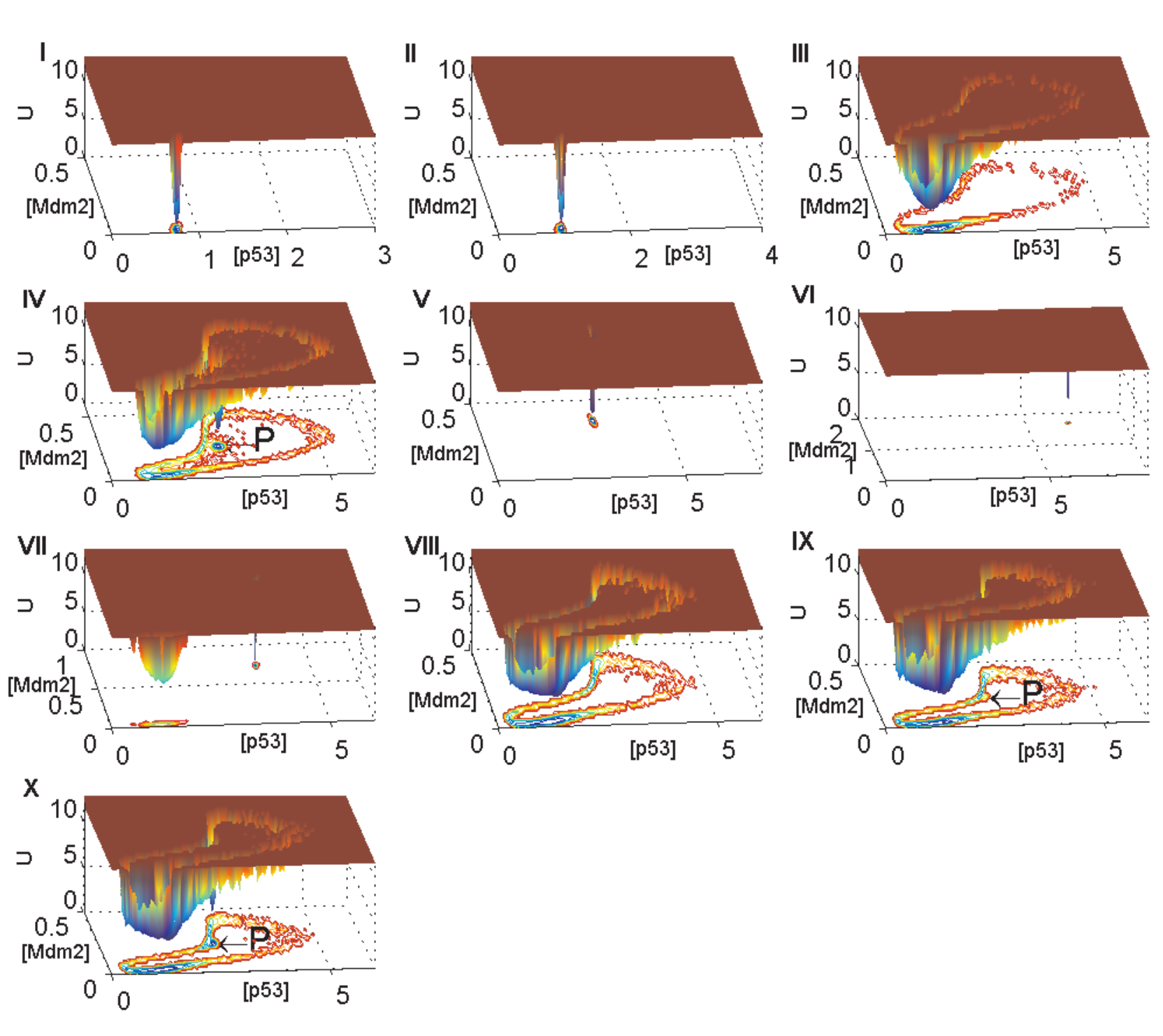}
\caption{The potential landscapes for the ten typical dynamics. Parameters are the same as in phase diagrams of Fig. \ref{fig:dyn}, and the diffusion matrix was taken as the diagonal matrix with noise strength $D=1.0 \times 10^{-6}$. Refer to the text for details.}
\label{fig:6}
\end{figure*}

To explore the global dynamics from the potential perspective, we projected the potential function to two independent variables, the p53 concentration $[\mathrm{p53}]$, and the total Mdm2 concentration $[\mathrm{Mdm2}]=[\mathrm{Mdm2}_{\mathrm{cyt}}]+[\mathrm{Mdm2}_{\mathrm{nuc}}]$. The potential landscapes for parameter values taken from the 10 typical regions are shown at Fig. \ref{fig:6}.

The potential landscapes show that when there is a single stable steady state (\textit{I}, \textit{II}, \textit{V}, \textit{VI}), the potential is funnelled towards a global minimum which corresponds to the global stable steady state.

In region \textit{VII}, there are two stable steady states, and the potential has two local minimum, corresponding to high or low p53 levels, respectively. The landscape at low p53 state has wide attractive region and shallow slop, while at the high p53 state has small attractive region and deep slope. These suggest that a cell from randomly select initial state is more likely to response with low p53 state because of the wider attractive region.

In other regions with p53/Mdm2 oscillations (\textit{III}, \textit{IV}, \textit{VIII}, \textit{IX}, \textit{X}), the potential landscapes show an irregular and inhomogeneous closed ring valley that corresponds to the deterministic stable limit cycle trajectory. It is obvious that the landscape is not uniformly distributed along the limit cycle path due to the inhomogeneous speed on the limit cycle \cite{wang2008potential,li2011landscape}. We also note that in the landscapes for regions \textit{IV}, \textit{IX} and \textit{X}, in addition to the closed valley, there is a deep funnel towards a local minimum. These local minimum show the coexistence of the stable steady state (marked by $P$ and the arrows in Fig. \ref{fig:6}).

\begin{figure}[htbp]
\centering
\includegraphics[width=8cm]{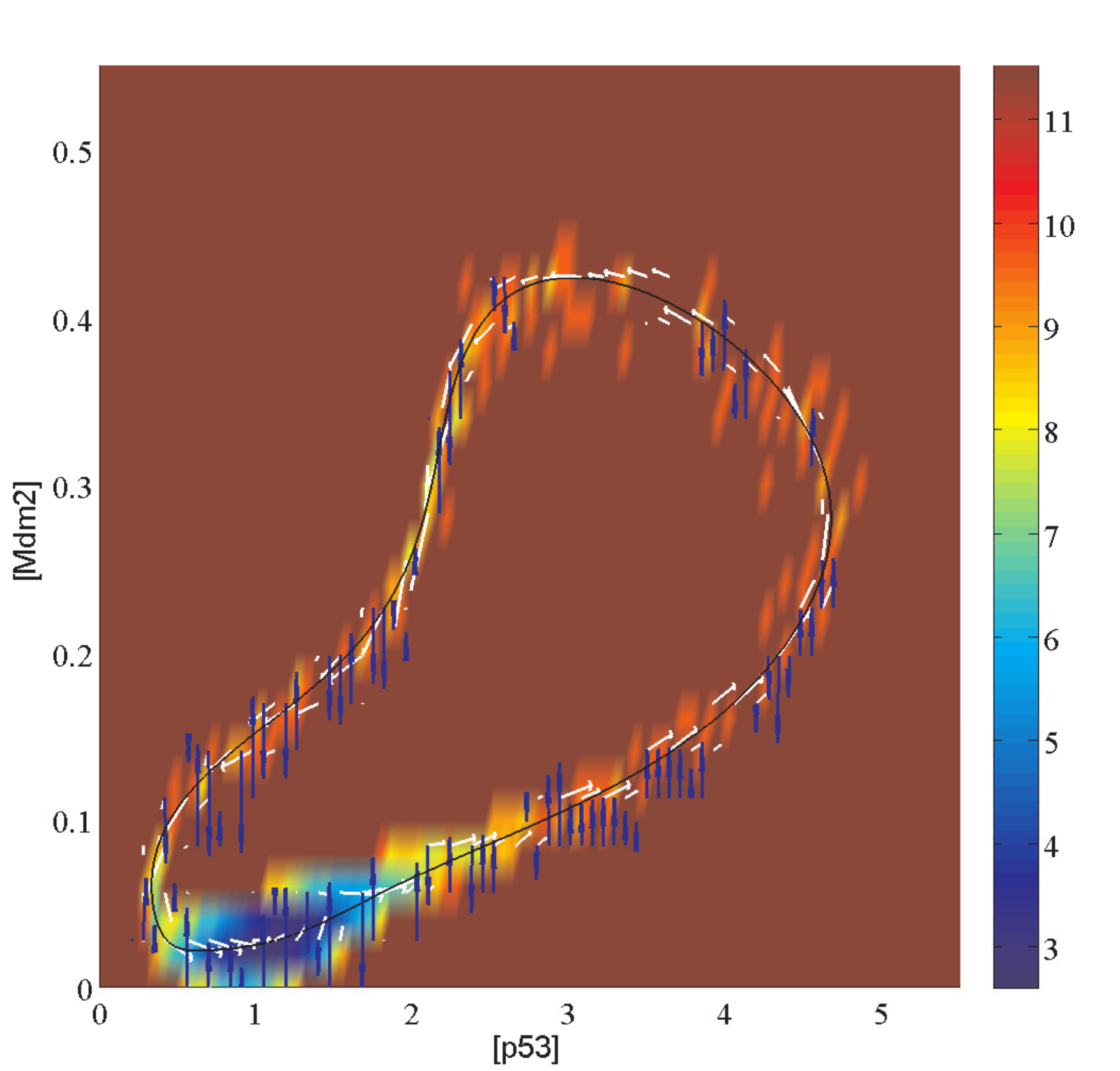}
\caption{Heat map of the potential landscape, superpose with  the potential force $-\dfrac{\partial\ }{\partial \mathbf{X}}U$ (blue arrows), stationary flux $\mathbf{J}_{ss}$ (white arrows) and the deterministic oscillation trajectory(black line). Here $\bar{v}_{\mathrm{p53}}=0.85, P_{0}=0.8$, and the noise strength $D=1.0\times 10^{-6}$. }
\label{fig:7}
\end{figure}

The regions with p53 oscillations are the most interested because the oscillation dynamics are essential for cell fate decision in response to various stresses. To further analyze the potential landscape when the system display stable p53 oscillations, we examined the case at region \textit{III} by calculating the potential force ($-\nabla U$) and the probability flux ($\mathbf{J}_{ss}$) (see Fig. \ref{fig:7}). As shown at Fig. \ref{fig:7}, the potential has local minimum valley along the deterministic oscillation trajectory. The values of potential are not uniformly distributed along the cycle. There is a global minimal potential at a state around $([\mathrm{p53},[\mathrm{Mdm2}])=(1, 0.02)$, which with small curl flux forces are large potential forces. The states with larger $[\mathrm{p53}]$ or $[\mathrm{Mdm2}]$ have higher potential, larger curl flux forces and smaller potential forces. We note that the potential forces are mostly vertical to the cycle path, and the curl flux forces are along the cycle. These results suggest that potential forces tend to attract the system states to the oscillation trajectory, while the curl flux forces drive the oscillation along the cyclical trajectory. The obvious low potential value at the global minimum indicates a pulse dynamics in each cycle the system states mostly stay at the state of global minimum while go through the other part of the limit cycle quickly. This pulse dynamics is in agree with previous studies in \cite{zhang2009cell,zhuge2014pdcd5}. 

\subsection {Effects of PDCD5 regulation efficiency}

PDCD5 interacts with the p53 pathway by promoting p53 stability through the disruption of the reaction between p53 and $\mathrm{Mdm2}_{\mathrm{nuc}}$. In our model, the p53 degradation rate function is given by a Hill-type function and the half-maximal effective concentration (EC50) is dependent on PDCD5 concentration given by Eq. (\ref{eq:K2}). In this function, the efficiency of PDCD5 is represented by the coefficient $\alpha_{1}$, larger $\alpha_{1}$ indicates higher efficiency. To investigate the effects of PDCD5 regulation efficiency, we took $(\bar{v}_{\mathrm{p53}}, P_0) = (0.85, 0.8)$ and changed $\alpha_{1}$ from zero to triple the default value to examine its effect on system dynamics.

\begin{figure}[htbp]
\centering
\includegraphics[width=8cm]{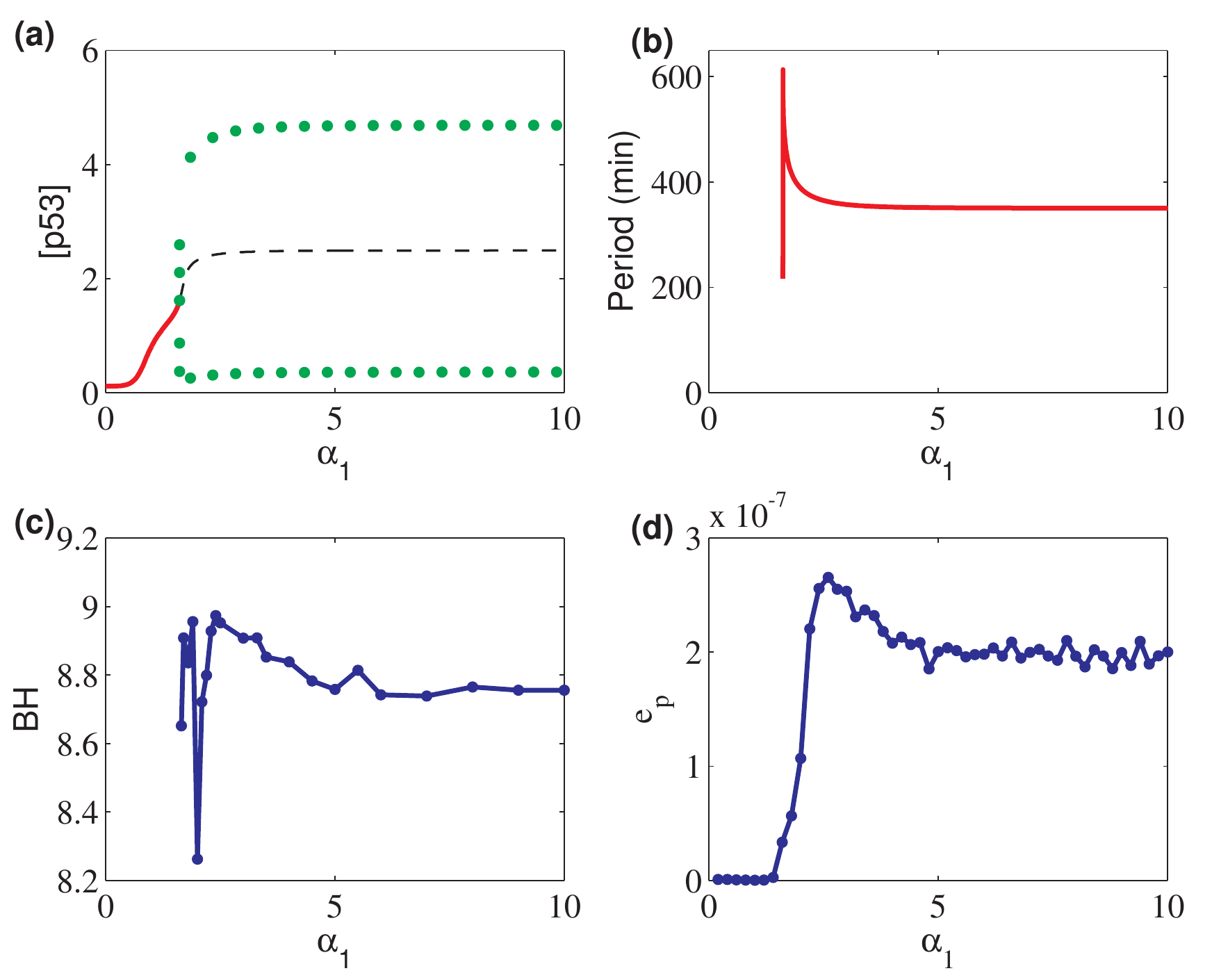}
\caption{Effects of PDCD5 regulation efficiency. (a) Bifurcation diagram with $\alpha_{1}$. Here $H$ marks the supercritical Hopf bifurcation point, red solid lines represent stable steady states, black dashed lines are unstable steady states, green solid dots are the maxima and minima of stable limit cycles. (b) Period  of stable limit cycles.  (c) The barrier height of stable limit cycles (BH). (d) The energy dissipation ($e_{p}$). Here $\bar{v}_{\mathrm{p53}}=0.85, P_{0}=0.8$, and the noise strength $D=1.0\times 10^{-6}$. }
\label{fig:8}
\end{figure}

Fig. \ref{fig:8}a shows the bifurcation diagram of p53 with respect to the efficiency $\alpha_{1}$. There is a supercritical Hopf bifurcation ($\alpha_1 = 1.612$) as $\alpha_{1}$ increases, where a stable limit cycle appears to yield p53 oscillations. The oscillation amplitude is insensitive with $\alpha_1$. The period increase rapidly near the Hopf bifurcation transition, and then decreases to $350 \mathrm{min}$ when $\alpha_1$ increases away the bifurcation point (Fig. \ref{fig:8}b).

To further study the stability of p53 oscillation, we calculated the barrier height in the potential landscape, which is a good measure of the stability of limit cycles \cite{wang2008potential}. The barrier height (BH) is defined by the potential function as $\mathrm{BH}=U_\mathrm{max}-U_\mathrm{min}$, where  $U_\mathrm{max}$ is the maximum potential inside the stable limit cycle, and $U_\mathrm{min}$ is the minimum potential along the cycle \cite{wang2008potential}. Higher barrier height means more stable. Fig. \ref{fig:8}c shows the barrier height versus $\alpha_{1}$. In our simulation, the barrier height is highly fluctuated near the Hopf bifurcation ($1.612<\alpha_1<2.6$), and then decreases towards a value of about $8.78$.

Next, we considered the effect of $\alpha_{1}$ to the dissipation of energy, which measures the inevitability for an open nonequilibrium system due to exchanging information and energies with its surrounding. The mean rate of energy dissipation ($h_{d}$) is give by $h_{d}=\int \mathbf{F}(\mathbf{X})\cdot \mathbf{J}(\mathbf{X},t)d \mathbf{X}$,  where $\mathbf{F}$ is the driving force, and $\mathbf{J}$ the probability flux vector. According to \cite{qian2006open}, the energy dissipation is equal to the entropy production rate ($e_\mathrm{p}$) at equilibrium state. Therefore, the energy dissipation can be used to obtain a global physical characterization of the nonequilibrium systems. Fig. \ref{fig:8}d shows the dissipation versus $\alpha_{1}$. The dissipation equals zero when $\alpha_{1}$ is smaller than the Hopf bifurcation value and there is a single steady state. This shows the detail balance at equilibrium state, in consistent with the fact that the least dissipation cost gives a more stable system in general \cite{han2008least}. When $\alpha_{1}$ increases beyond the Hopf bifurcation ($\alpha_1 > 1.612$), the energy dissipation increases rapidly to maximum at $\alpha_{1}=2.6$, and then decreases to and sustain at $2\times10^{-7}$ with $\alpha_{1}$ increasing.

In summary, these results show clearly transition dynamics when the system switches from a single stable steady state to a stable limit cycle with the increasing of PDCD5 efficiency. During the transition, the steady state is destabilized to yield p53 oscillations, both oscillation periods and energy dissipations first increase rapidly and then decrease to a sustain level, and the barrier height shows large fluctuation in the transition region. After the PDCD5 efficiency increases beyond the transition region, the system dynamics is insensitive to the efficiency.

\section{Conclusions}

The dynamics of tumor suppressor p53 plays important roles in the regulation of cell fate decision in response to various stresses\cite{lane1992cancer}. PDCD5 is known to interact with the p53 pathway and functions as a co-activator in p53 regulations. In p53 pathways, the p53-Mdm2 oscillator is crucial for cell response to DNA damage. In this study, we have systematically studied effects of PDCD5 to the p53-Mdm2 oscillator by methods of bifurcation analyses and potential landscapes. Our results reveal that the p53-Mdm2 oscillator can display monostability and bistability under low PDCD5 expression. When PDCD5 level is upregulated (for example, by DNA damage), p53 oscillations emerge by Hopf bifurcation, and bistability with the coexistence of both stable oscillation and a stable steady state is possible for proper PDCD5 level and p53 synthesis rate. These results were further verified by the potential landscapes, which clearly show the transition of landscapes with changing parameter values. We have also investigate the effects of PDCD5 efficiency in the interaction with p53 pathway, we showed that p53 oscillations can only be induced only when the efficiency is larger than a critical values of Hopf bifurcation, and the system dynamics show clear transition features in both barrier height and energy dissipation when the efficiency is close to the bifurcation point. Such abnormal behaviours at the transition region have been highlighted in recent years in the application of prediction and early diagnosis of complex diseases \cite{Liu:2013,Yu:2013,Zeng:2013}.

In this study we have focused on the PDCD5-regulated p53 dynamics in the p53-Mdm2 oscillator. Our results reveal global p53 dynamics mediated by PDCD5 and the levels of p53 production. However, a more complete p53 pathway is certainly necessary in understanding cell fate decisions in response to DNA damage. Further consideration of the effects of PDCD5 on a complete p53 network is certainly required in future studies.

\section*{Acknowledgements}

This work is supported by the National Natural Science Foundation of China (Nos. 91430101, 11272169, 11372017) and the scientific research project of Inner Mongolia colleges and universities (No. NJZY14130 ).

%\bibliography{byhpaper}

%\input{byhpaper.bbl}


\end{document}